# LGTBIDS: Layer-wise Graph Theory Based Intrusion Detection System in Beyond 5G

Misbah Shafi, *Student Member, IEEE*, Rakesh Kumar Jha, *Senior Member, IEEE,* Sanjeev Jain, *Senior Member, IEEE*

*Abstract*— The advancement in wireless communication technologies is becoming more demanding and pervasive. One of the fundamental parameters that limit the efficiency of the network are the security challenges. The communication network is vulnerable to security attacks such as spoofing attacks and signal strength attacks. Intrusion detection signifies a central approach to ensuring the security of the communication network. In this paper, an Intrusion Detection System based on the framework of graph theory is proposed. A Layerwise Graph Theory-Based Intrusion Detection System (LGTBIDS) algorithm is designed to detect the attacked node. The algorithm performs the layer-wise analysis to extract the vulnerable nodes and ultimately the attacked node(s). For each layer, every node is scanned for the possibility of susceptible node(s). The strategy of the IDS is based on the analysis of energy efficiency and secrecy rate. The nodes with the energy efficiency and secrecy rate beyond the range of upper and lower thresholds are detected as the nodes under attack. Further, detected node(s) are transmitted with a random sequence of bits followed by the process of re-authentication. The obtained results validate the better performance, low time computations, and low complexity. Finally, the proposed approach is compared with the conventional solution of intrusion detection.

## I. INTRODUCTION

One of the fundamental aspect of the existing and next generation communication network is the absolute fulfillment of network security. Security solutions are in the form of Firewall, Antivirus, and IDS (Intrusion Detection System). Firewall and antivirus are limited to the availability of states and their respective knowledge of the receiving host. IDS is a tool that scans, alerts, and detects malicious, unauthenticated entrances in the communication network [1]-[3]. Various advances were made in the technology of IDS to improve security solutions. Based on the technology, the IDS is categorized into three types, viz. anomaly-based IDS, misuse-based IDS, and hybrid IDS [4]. Anomaly-based IDS is further divided into three categories which are, data mining methodology, machine learning methodology, and statistical methodology [5]. There are several limitations associated with each of the adapted IDS mechanisms. These obstacles are mainly concerned with the volume of the data, labeled datasets, low accuracy, diversity, dynamics, low-frequency attacks, processing, and adaptability [6],[7]. At present, deep learning and machine learning-based intrusion detection mechanisms are rapidly developing. The major loopholes of these intelligent intrusion techniques include dependency on the characteristics of the sample data, poor robustness, and lower accuracy [8]. Therefore, an advanced IDS is required to be developed that can address these challenging measures.

### A. Related Work

In the literature, recent studies were focused on IDS to solve the problem of attacks in 5G (Fifth Generation), 5G NR (New Radio) and beyond. Pilot contamination attack [9] is addressed by the IDS involving the peak estimation algorithm. This algorithm is based on the technique of machine learning. For a dynamic scenario, the virtual channel peaks decide the normal state and contamination state. Pollution attack [10] 5G WCN (Wireless Communication Network) is addressed by the null space-based homomorphic message authentication code scheme. It enables to drop off the contaminated packets to make the network devoid of corrupted packet distribution. The scheme follows the random linear network coding mechanism. However, the mechanism does not involve the complete exclusion of the attacker(s) from access to the network. The Bandwidth spoofing attack [11] is another challenge in the multistage 5G WCN. The IDS for this attack model involves the probability analysis based on the power levels of the base station, relay, and SCA (Small Cell Access point). The approach of IDS is adept at detecting and eliminating the attacker(s) from the WCN. The security of the wireless sensor network is improved by the approach of the IDS based on the technique of deep learning [12]. The IDS based on artificial intelligence for the software-defined internet of things network is determined as an efficient approach. It is attained by the mechanism of HMM (Hidden Markov Model) detection of attack(s). The IDS mechanism involves the Bat algorithm for the optimal selection of features and a random forest scheme to design the classification methodology [13]. IDS involving neural networks is another efficient tactic to improve the security of the 5G WCN. The adaptive neuro-fuzzy inference system based on KDD 99 data set is utilized to detect the attack at the relay node of the 5G WCN [14].

### B. Motivation

Despite the advancement in WCN IDS technology, the complete security mechanism is still a challenge to be attained. The machine learning-based IDS involves 80% feature engineering efforts. The possibility of human errors during the process of feature extraction in such mechanisms is entirely evolving in nature [15]. The machine learning-based IDS is also associated with the class imbalance issue (elephant traffic and mouse traffic problem) and non-identical distribution issues

Misbah Shafi is with the department of electronics and communication, Shri Mata Vaishno Devi University, Katra, India 182230, Rakesh Kumar Jha is with the Indian Institute of Information Technology Design and Manufacturing Jabalpur, India 482005, and Sanjeev Jain is with the Central University of Jammu, Jammu and Kashmir, India 181143. (email: misbahshafi0@gmailcom, jharakesh.45@gmail.com, dr_sanjeevjain@yahoo.com)

TABLE I
COMPARISON OF LBGIDS WITH THE EXISTING INTRUSION DETECTION SCHEMES

| Ref. | Algorithm/scheme | Target technology | Target attack | Technique | Performance indicators | Optimized parameters | Network type |
|---|---|---|---|---|---|---|---|
| [9] | Peak estimation algorithm | 5G | Pilot contamination | Machine learning | Accuracy, DR, FAR, SNR | DA = 92.1 (for multiple attackers) DA = 96.8 (for single attacker), DR = 95%, FAR= 0.001% | Centralized |
| [10] | Null space-based homomorphic message authentication code (MAC) scheme | 5G | Pollution attack | random linear network coding (RLNC) | CO, communication overhead, decoding probability | CO = 0.005 to 0.046(seconds), communication overhead= 0.0051 to 0.0425, unsuccessful decoding probability = ~0.001(byzantine fabrication attack), ~0.005 (byzantine modification attack) | Distributed (Small cell network) |
| [11] | HMM (Hidden Markov Model) | 5G | Bandwidth spoofing attack | Machine learning (Prisoner's dilemma) | Transition probabilities | Probability of the valid user = 0.78 to 0.99, Probability of the intruder = 0.17 to 0.199 | Centralized |
| [12] | Restricted Boltzmann Machine (RBM) | Sensor networks | DoS, buffer overflow, portsweep, R2L | Deep learning | Detection rate, accuracy, False negative rate, F1 score, ROC | DR (%) = 95 (10 nodes) Accuracy(%)=96 (15 nodes) | Distributed network |
| [14] | Adaptive neuro fuzzy inference scheme | 5G | DoS (Land, Neptune, smurf, ping of death, tear drop) | Neural network | Average testing error, destination bytes, logged in bytes | Average testing error = 0.12 | Distributed |
| Proposed | LBGIDS algorithm | 5G and 5G beyond | Signal strength attacks, Half duplex attack, ping of death attack, DDoS | Graph theory | Computational time, detection rate, FAR, complexity, energy efficiency | Computational time =0.87s , DR(%)=98.3 , Accuracy (%) 92.2= , FAR = 1.854 | Distributed |

*DoS- Denial of Service, DDoS-Distributed Denial of Service, FAR- False Alarm Rate, DA- Detection Accuracy, DR-Detection Rate, FPR- False Positive Rate, Acc- Accuracy, CO-Computational Overhead, SNR- Signal to Noise Ratio,

between training data and test data. There are two main problems associated with deep learning-based IDS. The first problem is the requirement of a large amount of data, and the second is the problem of complex processing. To countermeasure, the limitations of the existing IDSs, an efficient and appropriate mechanism is required to improve the security of the WCN. The summary of the related work is given in Table I.

The graph theory is identified as a proficient methodology to emphasize the algorithmic and computational aspects to solve the problems in WCN. In [16], the concept of graph theory is applied to predict the channel modeling at the frequency band of 60 GHz. The quality of experience is improved by the operation of graph theory in the technologies of Device to Device (D2D) communication [17] beam scheduling [18], and cyber network configuration [19]. Also, the transmission order optimization problem in the case of D2D communication technology under TDD (Time Division Duplexing) cellular network is solved by the operation of the graph theory [20].

*C. Novelty*

Considering the graph theory as a proficient approach, an IDS is proposed in this paper based on the layer-wise analysis for the detection of the resource attack(s). To the extent of our knowledge, the concept of graph theory in view of security is not addressed by any of the existing IDS mechanisms. Also, layer-wise analysis incorporated in IDS is a novel approach to increase the efficiency of the IDS. The significant contribution of this paper is outlined as follows:

- Graph theory-based intrusion detection system has been proposed using layerwise execution to reduce complexity.
- The vulnerable nodes are achieved from each layer to perform the process of intrusion detection. The intrusion detection mechanism involves the investigation of the five recent attacks to provide high efficiency.
- The obtained simulation results verify that the proposed scheme is more efficient with reduced computational time and efficient security evaluations.

*D. Organization*

The paper is organized as follows. Section I presents the introduction. Section II describes the system model and problem formulation. The proposed mechanism of intrusion detection is discussed in Section III. Section IV provides the realization and representation of the proposed technique. The performance evaluation is analyzed in Section V. To the closure of the paper the conclusion is given in Section VI.

II. SYSTEM MODEL AND PROBLEM FORMULATION

The system model for the analysis of intrusion detection using the proposed IDS is presented in this section. Further, the problem formulation is illustrated considering LGTBIDS.

*A. System Model*

A WCN with multiple nodes with the variable configuration of each node is taken into account. The cell is configured by the layer-wise architecture based on the connection between the nodes.

Consider the communication scenario with $n, n \in \mathbb{N}$ number of nodes interacting with each other. The WCN is divided into $l$ number of layers. The layers are configured by the nodes followed by the immediate connections between the nodes. The base station forms the first layer and is present at the center of the cell. The immediate connections from the base station to the $a$ number of nodes form the second layer. Similarly, nodes forming the immediate connections from node to node define the layer-wise architecture. The second layer consists of $a$ number of nodes. The third layer consists of $b$ number of nodes. The cell edge nodes define the $l^{th}$ (last) layer of the communication network. The $l^{th}$ layer consists of $r$ number of nodes. Each layer is defined by the specific number of nodes participating in the communication network. For the mobile users in the network, the number of nodes in each layer may change with time frequently. In general, the in-between nodes form the layers to establish the communication from the first layer to the cell edge user. The set of layers is defined as $L = \{1, 2, 3, \ldots, l\}$. The first layer always consists of a single node and is quantified as $L(1) = \{1\}$. The set of nodes in the second layer is specified as $L(2) = \{1, 2, 3, \ldots, a\}$. The third layer consists of a set of nodes as $L(3) = \{1, 2, 3, \ldots, b\}$. Similarly, the $l^{th}$ layer contains a set of nodes as $L(l) = \{1, 2, 3, \ldots, r\}$.

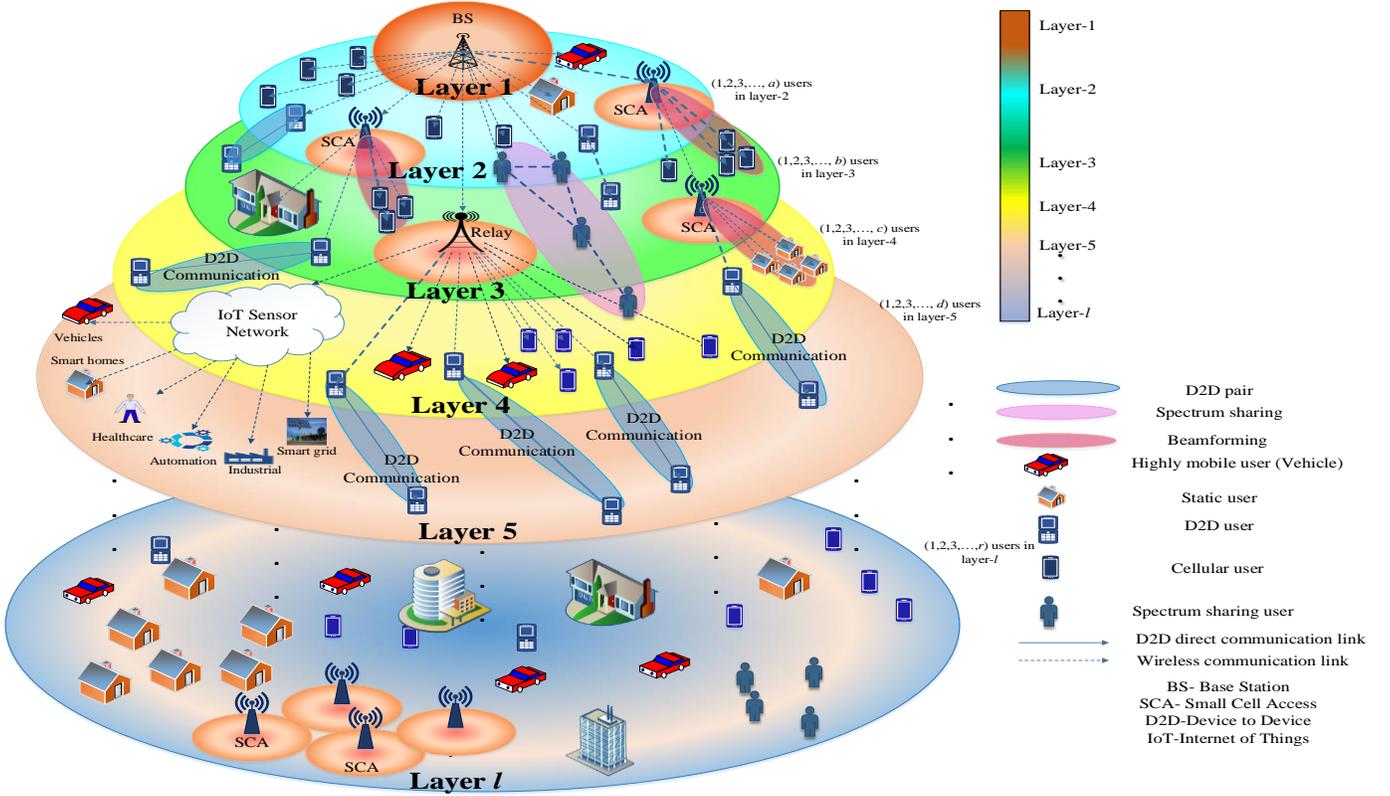

Fig. 1. System model for proposed LGTBIDS.

*B. Channel Modeling*

The layer-wise channel modeling is performed in the given architecture. An OFDM scheme is operated with the effect of Rayleigh flat fading. The signal $s_{i,m}^{j,k}$ is directed from the source node present in the source layer to the destination node present in the next layer. $i$ denotes the source node, $m$ is the respective layer of the source node, $j$ is the destination node, and $k \geq 2$ is the respective layer of the destination node. Where $i, m, j, k \in \mathbb{N}$. The received signal at the destination node is given by [26]:

$$D_{j,k}(i,m) = \sqrt{P_t^{j,k}(i,m)} h_{i,m}^{j,k} s_{i,m}^{j,k} + q_{i,m}^{j,k} \quad (1)$$

where $P_t^{j,k}(i,m)$ is the transmission power of the source node, $h_{i,m}^{j,k}$ is the channel coefficient present between source node $i$ and the destination node $j$.

$q_{i,m}^{j,k}$ is the AWGN (Additive White Gaussian Noise) with circular symmetry having zero mean and $\sigma_p^2$ as variance, such that $q_{i,m}^{j,k} \sim \text{CN}\left(0, \left(\sigma_p^2\right)_{i,m}^{j,k}\right)$. From equation (1), the SNR can be obtained as [26]:

$$\rho_{i,m}^{j,k} = \frac{P_t^{j,k}(i,m)\left|h_{i,m}^{j,k}\right|^2}{\left(\sigma_p^2\right)_{i,m}^{j,k}} \quad (2)$$

The equation (2) can be re-written as:

$$\rho_{i,m}^{j,k} = \frac{P_t^{j,k}(i,m)(|r_e|^2)_{i,m}^{j,k}\left(x_{i,m}^{j,k}\right)^{-\alpha}}{\left(\sigma_p^2\right)_{i,m}^{j,k}} \quad (3)$$

where $|r_e|^2$ denotes the Rayleigh fading channel gain, $\alpha$ is the path coefficient. $x_{i,m}^{j,k}$ is the distance between source node $i$ and the destination node $j$. The capacity using Shannon's capacity while incorporating $\beta$ as the bandwidth is given by [27],[28]:

$$C_{i,m}^{j,k} = \beta \log_2\left(1 + \rho_{i,m}^{j,k}\right) \quad (4)$$

The parameter of secrecy rate is analyzed to observe the intruder in the network. The secrecy rate is given by:

$$(C_s)_{i,m}^{j,k} = \begin{cases} \left|C_{i,m}^{j,k} - C_{i,m}^e\right| & C_{i,m}^{j,k} \geq C_{i,m}^e \\ 0 & o.w \end{cases} \quad (5)$$

*Proposition 1:* The communication network is defined by the graph theory such that each node represents the vertex. The edge is represented as the wireless communication link between the nodes specified by the parameter of estimated capacity. The detection of the attack is uniquely identified by examining the vertices covering the nodes in the form of secrecy rates and energy efficiency. (Refer to Fig. 1, Fig. 2.)

TABLE II
TABLE OF SYMBOLS

| Symbol | Description |
| --- | --- |
| $i, m$ | Source node, layer of the source node |
| $j, k$ | Destination node, layer of destination node |
| $V_s^c$ | Lower bound secrecy rate |
| $V_s$ | Upper bound secrecy rate |
| $V_s^{ach}$ | Achieved secrecy rate |
| $EE$ | Upper bound energy efficiency |
| $EE^c$ | Lower bound energy efficiency |
| $EE^{ach}$ | Achieved energy efficiency |
| $f(x)$ | Range cross detection function |
| $n$ | Number of nodes in the network |
| $l$ | Last layer in the communication network |
| $E^e$ | Intruder capacity |
| $P_e$ | Additional propagation loss |
| $E^k$ | Capacity of the vertex |
| $E$ | Minimum capacity vertices matrix |

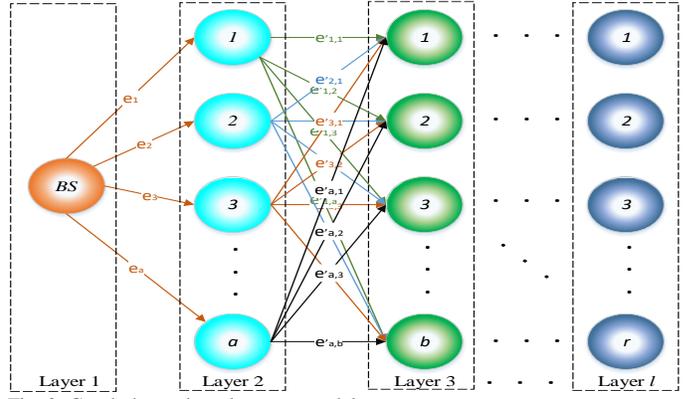

Fig. 2. Graph theory based system model.

The equation (5) can also be formulated using the concept of the graph as:

$$(V_s)_{i,m}^{j,k} = |E_{i,m}^{j,k} - E_{i,m}^e| \quad E_{i,m}^{j,k} \geq E_{i,m}^e \quad (6)$$

The energy efficiency (bps/watt) is stated as [31]:

$$EE_{i,m}^{j,k} = \frac{E_{i,m}^{j,k}}{P_{cc}} \quad (7)$$

Where $P_{cc}$ is the total power consumption in watts. The valid user must satisfy equations (6) and (7). The lower bound of the secrecy rate and energy efficiency is obtained by the inclusion of additional propagation loss. The additional propagation loss is due to the presence of hydrometeors or dust. The signal attenuation while propagating through different $z$ entities of depth $w$ is approximated as [25]:

$$(P_e) = e^{-\gamma w} \quad (8)$$

Where $\gamma$ is the constant of attenuation and is dependent on the dielectric properties and material of the entity. Considering equal approximation of $\gamma$, equation (8) is approximated as:

$$(P_e) = e^{-\gamma(w_1 + w_2 + w_3 + \cdots + w_n)} \quad (9)$$

Where, $w_1, w_2, w_3, \ldots, w_n$ is the depth of the random entities covering the path of the signal. The channel coefficient (dB) while incorporating additional propagation loss (dB) is given by:

$$|h^c{}_{i,m}^{j,k}|^2 = -((P_l)_{i,m}^{j,k} + P_e) \ldots \ldots \ldots \ldots (10)$$

Where, $P_l$ is the free space path loss. Using equation (10) in equations (2) and (4), the lower bound secrecy rate is given by:

$$(V_s^c)_{i,m}^{j,k} = |(E^c)_{i,m}^{j,k} - E_{i,m}^e| \quad (11)$$

The lower bound of energy efficiency (bps/watt) is stated as:

$$(EE^c)_{i,m}^{j,k} = \frac{(E^c)_{i,m}^{j,k}}{P_{cc}} \quad (12)$$

Where, $E^c$ is the lower bound capacity while encompassing additional propagation loss. Similarly, the channel coefficient for the upper bound parameters is obtained as

$$|h_{i,m}^{j,k}|^2 = -((P_l)_{i,m}^{j,k}) \quad (13)$$

Using equation (13) in equations (2) and (4), the upper bound secrecy rate without the inclusion of any propagation losses as.

$$(V_s)_{i,m}^{j,k} = |(E)_{i,m}^{j,k} - E_{i,m}^e|$$

The upper bound of energy efficiency (bps/watt) is stated as:

$$(EE)_{i,m}^{j,k} = \frac{(E)_{i,m}^{j,k}}{P_{cc}} \quad (14)$$

Where $E$ is the upper bound capacity without additional propagation loss.

*Proposition 2:* To improve the efficiency of the intrusion detection system, the most vulnerable nodes from the network are attained first which are then followed by the analysis of intrusion detection.

C. Problem Formulation

The key aspect of our proposed mechanism is the detection of the intruder in the communication network with reduced complexity while maintaining accuracy. Initializing the parallel processing for each layer, the nodes with minimum capacity are defined by the edge matrix $E$ using equation (4) as

$$E = min(E_{i,m}^{j,k}) \quad (15)$$

Such that the detection process is defined for the nodes with the maximum chances of attack. Therefore equation (15) defines the matrix of vulnerable nodes. For the nodes specified in the matrix $E$ of equation (15), the detection of the malicious node(s) is defined by the first criterion of secrecy rate using equations (6) and (11) as

$$(V_s^c)_{i,m}^{j,k} \leq (V_s^{ach})_{i,m}^{j,k} \leq (V_s)_{i,m}^{j,k} \quad (16)$$

Similarly, for the nodes specified in equation (15), the second criterion of energy efficiency using equations (7) and (12) is given by

$$(EE^c)_{i,m}^{j,k} \leq (EE^{ach})_{i,m}^{j,k} \leq EE_{i,m}^{j,k} \quad (17)$$

Constraint in equations (16) and (17) ensures the detection of the malicious nodes in the communication network. The expression in equations (16) and (17) are defined for those nodes only that are obtained from equation (15), thereby reducing the burden of evaluating the secrecy rate for all the nodes in the network. Moreover, the list of possible vulnerable nodes can be attained in equation (15). (Refer to Lemma 1) ∎
To improve the security performance of the network, the resultant node(s) is re-authenticated and the edge with a greater secrecy rate is allocated. Till then the detected node is transmitted with the random sequence of bits. In case, if re-authentication fails then the detected node is continuously transmitted with the random sequence of bits.

## III. LGTBIDS Algorithm: How it Works

### A. Attack Description

In bandwidth spoofing attack, the bandwidth is spoofed from the valid user and is allocated to the intruder. The spoofing of the bandwidth by the attacker is obtained by jamming the network to an extent such that the spoofing takes place [11].

In half-duplex attack, spoofing the resource is the primary goal. The target user is attacked at the downlink only during the process of resource allocation, such that the resources that are meant to be allocated to the legitimate user are spoofed and allocated to the attacker. The attack involves the identity spoofing, jamming, and resource spoofing of the valid user. The attack possibility is identified for the 5G and beyond 5G wireless communication networks [25].

In UAV attack, the intruder masquerades as the valid SCA (Small Cell Access) point such that the targeted valid user is connected with the intruder SCA instead of the valid SCA [27].

In handover attack, the attacker targets the high-speed user in the handover zone. The attack can result in the hijacking, jamming, spoofing, or masquerading of the valid user [28].

In DoS (Denial of service), the target user is flooded with unnecessary services thereby disrupting the network and making the valid user unavailable for the intended communication [29].

### B. Detection Model

The majority of threats in the communication network incorporate the attacks based on the deteriorated secrecy rate and CSI (Channel State Information). In half-duplex attack, handover attack, UAV attack DoS (Denial of service), and bandwidth spoofing attack, the primary parameters that are drastically affected include capacity, secrecy rate, and energy efficiency.

The proposed algorithm is identified as a novel approach to eliminate such attacks. The proposed approach is defined as a layer-based methodology to predict the detection of attacks at the nodes of the communication network using graph theory. Moreover, it provides the adaptability to perform the comparative analysis of the respective vulnerable node. The proposed methodology involves four phases for the prediction of intrusion detection with the procedure of improvement in security performance.

*1) Phase of graph plot formation*

This phase is the simplest of all the phases. The whole communication network is implied in the form of vertices and edges. The vertex can be the representation of each user(node) in the network. The communication link is specified by the edges between the nodes. Fig. 2. depicts the communication scenario using graph theory. The graph is classified into layers based on the connection between nodes. The BS is considered as the first layer. The immediate directly connected vertices from the BS form the second layer. Similarly, the nodes directly connected from the second layer vertices form the third layer and so on up to the last layer where no further connections are made by the vertices.

*2) Phase of preliminary screening*

The phase involves the procedure of screening the layers to achieve vulnerable vertices. The screening phase is based on the process of obtaining vertices with minimum capacity, one vertex from each layer. As per the criteria of Wyner's theory of wiretap channels, the security of information is maintained iff the quality of the main legitimate channel is greater than the quality of the wiretap illegitimate channel. In other words, the attack is the possibility is maximum when the capacity of the valid user is less or equal to that of the capacity of the intruder. Considering the criteria, the vulnerable nodes are achieved based on the minimum capacity concept. The vertices are obtained where the possibility of the attack is maximum. It provides the referral vertices for the further analysis of intrusion detection in the network. This phase incorporates the initializing of parallel processing for each layer. The estimation is quite simple without the involvement of complex progressions.

*Illustration:* Consider $l$ number of layers formed in a communication network. First layer with $a$ number of vertices, second layer with $b$ number of vertices, and so on up to $l^{th}$ layer with $r$ number of vertices in the communication network. Using equation (4) the capacity of the vertices forming layers $1, 2, 3, \ldots, l$ is represented by a vector as

$$E^k = \left[E_{i,m}^{j,k}\right] \quad (18)$$

Where $k \geq 2$ as the BS forms the first layer, $k$ represents the destination layer. We consider the linear graph the vector representing the capacity of vertices in layer 2 is given by

$$E^2 = \left(e_{1,1}^{1,2}, e_{1,1}^{2,2}, e_{1,1}^{3,2}, \ldots, e_{1,1}^{a,2}\right) \quad (19)$$

For layer 3 the capacity vector is specified as

$$E^3 = \left(e_{1,2}^{1,3}, e_{1,2}^{2,3}, e_{1,2}^{3,3}, \ldots, e_{1,2}^{b,3}, \ e_{2,2}^{1,3}, e_{2,2}^{2,3}, e_{2,2}^{3,3}, \ldots, e_{2,2}^{b,3}, \right.$$
$$\left. e_{3,2}^{1,3}, e_{3,2}^{2,3}, e_{3,2}^{3,3}, \ldots, e_{3,2}^{b,3}, \ldots, \ e_{a,2}^{1,3}, e_{a,2}^{2,3}, e_{a,2}^{3,3}, \ldots, e_{a,2}^{b,3},\right) \quad (20)$$

In equation (20) we have assumed every vertex of layer 2 has a relation with every vertex of layer 3. Similarly, for $l^{th}$ layer the capacity is stated as

$$E^l = \left(e_{1,l-1}^{1,l}, e_{1,l-1}^{2,l}, e_{1,l-1}^{3,l}, \ldots, e_{1,l-1}^{r,l}, \ e_{2,l-1}^{1,l}, e_{2,l-1}^{2,l}, e_{2,l-1}^{3,l}, \ldots, e_{2,l-1}^{r,l}, \right.$$
$$\left. e_{3,l-1}^{1,l}, e_{3,l-1}^{2,l}, e_{3,l-1}^{3,l}, \ldots, e_{3,l-1}^{r,l}, \ldots, \ e_{\varphi,l-1}^{1,l}, e_{\varphi,l-1}^{2,l}, e_{\varphi,l-1}^{3,l}, \ldots, e_{\varphi,l-1}^{r,l},\right) \quad (21)$$

Where $\varphi \in \mathbb{N}$ and is the number of vertices present in the penultimate layer. The vertices with minimum capacity are obtained from each layer given by

$$E = \begin{bmatrix} \min(E^2) \\ \min(E^3) \\ \vdots \\ \min(E^l) \end{bmatrix}_{(l \times 1)} \quad (22)$$

The resultant matrix $E$ in equation (22) provides only vertices with minimum capacity from the corresponding layer. The resultant matrix is of the order of $(l \times 1)$. One vertex from each layer is analyzed as the choice of an attack.

*3) Phase of intrusion detection*

The phase performs the layer-wise process of intrusion detection on the acquired vertices from the phase of preliminary screening given in equation (22). Initially, the process starts from the estimation of upper and lower bounds of secrecy rate and energy efficiency for the vertices obtained from equation (22). The intrusion detection is further applied by using the range cross detection function. The range cross detection function compares the respective achieved secrecy rate and energy efficiency of the vertex with the respective range between estimated upper and lower bounds.

*Illustration:* From equation (22), the upper bound secrecy rate matrix for each vertex with minimum capacity from each layer is declared by using equation (6) as

$$V_s = \begin{bmatrix} (V_s)_{i,m}^{j,1} \\ (V_s)_{i,m}^{j,2} \\ \vdots \\ (V_s)_{i,m}^{j,l} \end{bmatrix}_{(l \times 1)} \quad (23)$$

From equation (22) the lower bound secrecy rate matrix for the deduced vertices by using equation (11) is shown as

$$V_s^c = \begin{bmatrix} (V_s^c)_{i,m}^{j,1} \\ (V_s^c)_{i,m}^{j,2} \\ \vdots \\ (V_s^c)_{i,m}^{j,l} \end{bmatrix}_{(l \times 1)} \quad (24)$$

Also, the upper bound $EE$ and lower bound $EE^c$ energy efficiency matrices for the obtained vertices by using equations (7) and (12) are interpreted as

$$EE = \begin{bmatrix} (EE)_{i,m}^{j,1} \\ (EE)_{i,m}^{j,2} \\ \vdots \\ (EE)_{i,m}^{j,l} \end{bmatrix}_{(l \times 1)} \quad \text{and} \quad EE^c = \begin{bmatrix} (EE^c)_{i,m}^{j,1} \\ (EE^c)_{i,m}^{j,2} \\ \vdots \\ (EE^c)_{i,m}^{j,l} \end{bmatrix}_{(l \times 1)} \quad (25)$$

In addition to it, the achieved secrecy rate matrix $V_s^{ach}$ and achieved energy efficiency matrix $EE^{ach}$ for the achieved vertices in equation (22) is given as

$$V_s^{ach} = \begin{bmatrix} (V_s^{ach})_{i,m}^{j,1} \\ (V_s^{ach})_{i,m}^{j,2} \\ \vdots \\ (V_s^{ach})_{i,m}^{j,l} \end{bmatrix}_{(l \times 1)} \quad (26)$$

$$EE^{ach} = \begin{bmatrix} (EE^{ach})_{i,m}^{j,1} \\ (EE^{ach})_{i,m}^{j,2} \\ \vdots \\ (EE^{ach})_{i,m}^{j,l} \end{bmatrix}_{(l \times 1)} \quad (27)$$

The range cross detection function is used to detect the intrusion detection and is stated as

$$f(x) = \begin{cases} 0 & f(V_s^{ach}) \in f(V), f(EE^{ach}) \in f(EE') \\ 1 & otherwise \end{cases} \quad (28)$$

Where $x = intruder\ detection$, $V$ lies between $V_s^c$ to $V_s$ and $EE'$ lies between $EE^c$ to $EE$. Two classes can be considered for the analysis. One is the intruder and the other is the valid user. Following the strategy of the two classes, the one-hot encoding mechanism is evaluated. 0 represents that the corresponding vertex is a valid user and 1 signifies the vertex as an intruder. Using equations (23)-(27) in equation (28) we get

$$A = f(x) = \begin{bmatrix} (y)_{i,m}^{j,1} \\ (y)_{i,m}^{j,2} \\ \vdots \\ (y)_{i,m}^{j,l} \end{bmatrix}_{(l \times 1)} \quad (29)$$

Where, $y = 0\ or\ 1$. The equation (29) performs the mapping function. The vertex that does not map within the range defined between the upper bound and lower bound produces the output as 1. The 1 in the output ultimately declares that the corresponding node is under intrusion.    (Refer Lemma 1)∎

*4) Phase of disconnection and re-authentication*
The detected vertex further undergoes the procedure of disconnection. Here disconnection means the communication link is kept silent. The silent link is defined as the transmission of a random sequence of bits. The detected vertex is requested to pass through the authentication again. If the re-authentication fails the procedure of silent link continues. Otherwise, the process continues till the edge with a greater secrecy rate is available and is allocated to the detected vertex for the prevention of the attack.

The advantages of layerwise processing of intrusion detection include the high performance of intrusion detection in a layered approach due to reduced complexity in network specification. Parallel processing of intrusion detection on layers is executed which reduces the time complexity. Instead of executing the intrusion detection mechanism on the whole network, layered execution simplifies the intrusion detection scenario. Easy update of a number of nodes in the communication network on the entry and exit of a node. Easy identification of the node under intrusion in layerwise approach. Maintenance of intrusion detection mechanism is easier in the layerwise phenomenon.

## IV. REALIZATION AND REPRESENTATION OF LGTBIDS

This section denotes the pseudo-code of the proposed LGTBIDS scheme. The pseudo-code provides a convenient understanding of the implementation and procedure of the intrusion detection mechanism in the next generation wireless communication networks. The pseudo-code consists of two parts. The first part describes the realization of the communication scenario using graph theory. The second part defines the process of intrusion detection.

The communication network is randomly distributed with $n$ the number of nodes such that $n \in \mathbb{N}$. The whole network is represented in the form of vertices and edges while incorporating graph theory. Based on the immediate connection between the vertices the layer-wise distribution takes place. Accordingly, the numeral of the layers is updated. BS is considered as the first layer in all communication networks. The graph plot mechanism is shown in pseudo-code 1.

In pseudo-code 2 the procedure of intrusion detection is operated. Initializing from the process of obtaining vulnerable vertices of the network. One possible vulnerable vertex is achieved from each layer. The BS is not included in the vector of vulnerable vertices. Further analysis of intrusion detection takes place on the achieved vulnerable vertices. The upper bound secrecy rate, lower bound secrecy rate, achieved secrecy rate, upper bound energy efficiency, lower bound energy efficiency, and the achieved energy efficiency of the corresponding vulnerable vertex are estimated. The mapping procedure is followed by using the range cross detection function. The function makes use of the upper and lower bounds and tends to acquire only the vertices under intrusion. To ameliorate the security performance of the network the detected nodes are transmitted with the random sequence of bits. In addition to that, for the detected vertex the edge within the range of upper and lower bounds is found out. Further followed by the re-authentication process from the server again. The respective vertex is safeguarded after completing the re-authentication process successfully. If the re-authentication fails, the detected vertex is continued with the transmission of a random sequence of bits, and the information of the detected vertex is transmitted to the server. The arrangement of the steps is concisely defined in pseudo code 2.

| **Pseudo Code 1 for realization of graph theory based communication scenario** |
|---|
| 1: **Input parameters** |
| Number of nodes in the network $n$, Path loss $p_l$ |
| Additional propagation loss $p_e$ |
| Transmission power $p_t$ |
| Operating bandwidth $\beta$, Noise power $\sigma_p$ |
| 2: **Initialization** |
| Initialize BS |
| Set of nodes $N = [1\ 2\ 3\ ...\ n]$ |
| Set of communication links $E = [1\ 2\ 3\ ...\ u]$ /*$(n,u) \in \mathbb{N}$*/ |
| 3: **Network under graph theory representation** |
| **for** $i = 1:n$ |
| $G = digraph\ (N, E)$ |
| Determine number of vertices in the connecting edges from the BS |
| Define layers $L = [1\ 2\ 3\ ...\ l]$ |
| **end for** |
| 4: **Updation in the number of nodes** |
| **if** $n \neq n$ |
| update $L$ |
| **end if** |

| **Pseudo Code 2 for LGTBIDS** |
|---|
| 1: **Loop statement for vulnerable vertices** |
| **Parfor** $i = 1:l$ |
| **for** $j = 1:r$ /*r-number of nodes in a layer */ |
| Compute SNR $\delta(i,j)$ |
| Compute capacity $E$ |
| $E(i,j) = \beta \log_2(1 + \delta(i,j))$ |
| **end for** |
| Store the result in the form of a vector $E(i) = (\bar{E})$ |
| Find the vulnerable nodes $E'(i) = [\min(E(i))]$ |
| 2: **Compute upper and lower bounds using equation (11),(12),(13) and (14) for the $E'$ vertices** |
| $V_s(i) = E(i) - E^e(i)$ /*general expression for secrecy rate*/ |
| $EE(i) = \frac{E(i)}{P_c(i)}$ /*general expression for energy efficiency*/ |
| 3: **Applying range cross detection function using equation (28)** |
| **if** $f(x(i)) = 1$ |
| Intruder is detected |
| Print intrusion detected nodes in $F$ |
| Transmit $A$ to $F$ |
| **elseif** $f(x(i)) = 0$ |
| user is valid |
| **endif** |
| **for** $g = 1: Range(F)$ |
| Find $E'(i,g) \in f(x(i,g)) = 0$ for time stamp (t)/*for sensitive networks like defence the time stamp is quite reduced*/ |
| **if** $E'(i,g) \in f(x(i,g)) \neq 0$ |
| Transmit $A$ to $F(i,g)$ /*A=random sequence of bits*/ |
| **else** (follow the authentication process from the server) |
| **end if end for end parfor** |
| **if** $n \neq n$ |
| go to step 4 of pseudo code 1 |
| **end if** |
| 4: **Plot the results** |

TABLE III
SIMULATION PARAMETERS

| Parameter | Value |
|---|---|
| Operating frequency | 28GHz |
| Bandwidth | 800MHz |
| Noise power | -106dBm |
| Cell radius | 250m |
| Number of layers | 5 |
| Number of vertices | 24(including BS, SCA, Relay) |
| Number of edges | 24 |
| User speed(mobile) | 60Kmph |
| BS power (max.) | 30dBm |
| BS power (min.) | 18dBm |
| Path loss exponent | 2 |
| Number of vertices in layer 2 | 5 |
| Number of vertices in layer 3 | 12 |
| Number of vertices in layer 4 | 4 |
| Number of vertices in layer 5 | 2 |

## V. RESULTS AND DISCUSSIONS

This section represents the performance evaluation of the proposed LGTBIDS scheme. The microcell scenario with a maximum radius of 250m is taken into consideration for the analysis of the proposed mechanism. The maximum of 5 possible layers is analyzed with the characterization of nodes in each layer. The simulation plot using LGTBIDS is shown in Fig. 3. However, more layers can be formed depending on the connection of the nodes present in the communication network. Other parameters that provide insight into the performance of the proposed mechanism include sensitivity, FNR (False Negative Rate), specificity, FOR (False Omission Rate), balanced accuracy, F1 score, and error rate. These characteristic features define the predictive performance of the proposed IDS. Sensitivity offers a fraction of nodes under attack that is predicted accurately. The higher the value of sensitivity higher is the possibility of detecting the attack. FNR defines the proportion of nodes under which are declared valid nodes. FOR defines the metric of false negatives detected incorrectly. The smaller the FOR value more is the accuracy of detecting valid nodes. Specificity declares the actual fraction of valid nodes that are predicted correctly. Consequently, the higher the value of specificity, the greater the performance of the network. Balanced accuracy is defined as the average of correct fractions viz sensitivity and specificity. F1 score is an accuracy measure of the detection system. Error rate defines the inaccuracy in the detected node(s) and determines the effectiveness of the system.

### A. Simulation Environment

The proposed intrusion detection algorithm is implemented in a MATLAB simulation environment. The parameters of standardization for different node types include SCA, relay, cellular user, spectrum sharing device, mobile nodes (Vehicles), and static nodes. A communication scenario with the layer-wise approach is followed to detect intruder detection in the network. The execution of the algorithm follows the basis of graph theory with the features of nodes and vertices. The proposed algorithm is devoid of the requirements including training procedures and high memory processor ends. The algorithm involves the technique of detection with regard to the characteristics of the node. The characteristics are defined by the parameter of achieved secrecy rate and energy efficiency. Table III represents the simulation parameters contemplated for the evaluation of the results [30]. Table IV represents the estimated calculations to provide insight into the illustration of LGTBIDS.

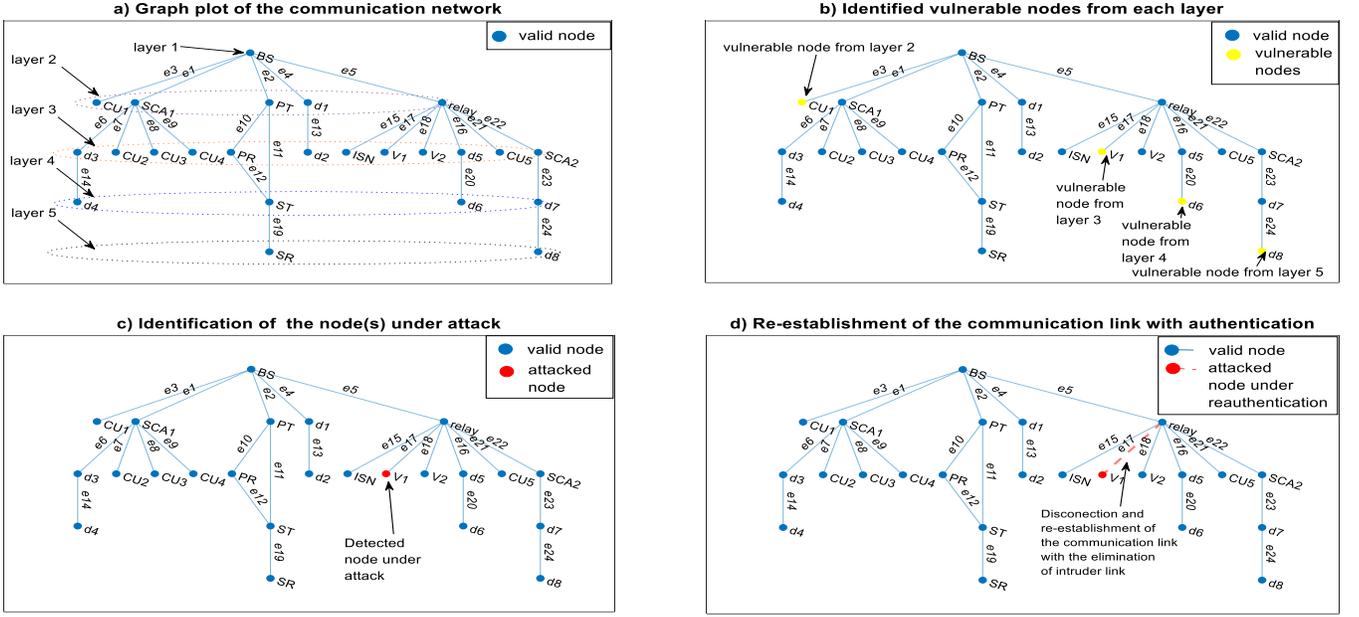

Fig. 3. LGTIDS plot based on graph theory for L=5

TABLE IV
LGTBIDS UPPER AND LOWER BOUND ILLUSTRATION

| Layer | Node Type | Maximum transmission power | Lower bound Secrecy Rate(Gbps) | Upper bound Secrecy Rate(Gbps) | Lower bound Energy Efficiency (Gbps/Watt) | Upper bound Energy Efficiency (Gbps/Watt) | Achieved Energy Efficiency (Gbps/Watt) | Achieved Secrecy Rate (Gbps) | Normal/Vulnerable/Under attack |
|---|---|---|---|---|---|---|---|---|---|
| Layer 2 | CU1 | 23dBm | 6.034 | 6.468 | 0.302 | 0.324 | 0.323 | 6.457 | **Vulnerable** |
| | SCA1 | 27dBm | 7.740 | 8.019 | 0.387 | 0.401 | 0.401 | 8.012 | Normal |
| | PT | 24dBm | 10.419 | 10.484 | 0.522 | 0.525 | 0.525 | 10.482 | Normal |
| | d1 | 24dBm | 9.835 | 9.835 | 0.492 | 0.492 | 0.498 | 9.950 | Normal |
| | Relay | 24dBm | 6.355 | 6.757 | 0.318 | 0.338 | 0.338 | 6.747 | Normal |
| Layer 3 | d3 | 24dBm | 7.812 | 8.479 | 15.588 | 16.916 | 16.884 | 8.462 | Normal |
| | CU2 | 23dBm | 8.509 | 9.084 | 16.978 | 18.126 | 18.098 | 9.070 | Normal |
| | CU3 | 23dBm | 10.522 | 10.882 | 20.994 | 21.713 | 21.696 | 10.873 | Normal |
| | CU4 | 23dBm | 7.583 | 8.282 | 15.131 | 16.525 | 16.490 | 8.264 | Normal |
| | PR | 24dBm | 7.989 | 8.499 | 31.804 | 33.837 | 33.787 | 8.486 | Normal |
| | d2 | 24dBm | 7.272 | 7.804 | 21.888 | 21.992 | 21.888 | 7.791 | Normal |
| | ISN | 15dBm-24dBm | 8.009 | 8.649 | 15.981 | 17.257 | 17.226 | 8.633 | Normal |
| | V1 | 13dBm | 6.173 | 7.095 | 12.317 | 14.157 | 12.272 | 6.150 | **Under attack** |
| | V2 | 13dBm | 6.482 | 7.352 | 12.934 | 14.669 | 14.626 | 7.330 | Normal |
| | d5 | 24dBm | 8.172 | 8.790 | 16.305 | 17.538 | 17.508 | 8.774 | Normal |
| | CU5 | 23dBm | 7.851 | 8.512 | 15.665 | 16.984 | 16.952 | 8.496 | Normal |
| | SCA2 | 27dBm | 8.231 | 8.790 | 16.423 | 17.538 | 17.511 | 8.776 | Normal |
| Layer 4 | d4 | 24dBm | 3.316 | 4.457 | 26.347 | 35.408 | 35.181 | 4.429 | Normal |
| | ST | 24dBm | 6.194 | 6.940 | 24.662 | 27.628 | 27.555 | 6.921 | Normal |
| | d6 | 24dBm | 3.130 | 4.305 | 24.864 | 34.199 | 33.964 | 4.275 | **Vulnerable** |
| | d7 | 24dBm | 6.024 | 6.972 | 12.019 | 13.912 | 13.865 | 6.949 | Normal |
| Layer 5 | SR | 24dBm | 4.072 | 5.180 | 16.211 | 20.623 | 20.513 | 5.152 | Normal |
| | d8 | 24dBm | 3.704 | 4.883 | 29.422 | 38.792 | 38.557 | 4.854 | **Vulnerable** |

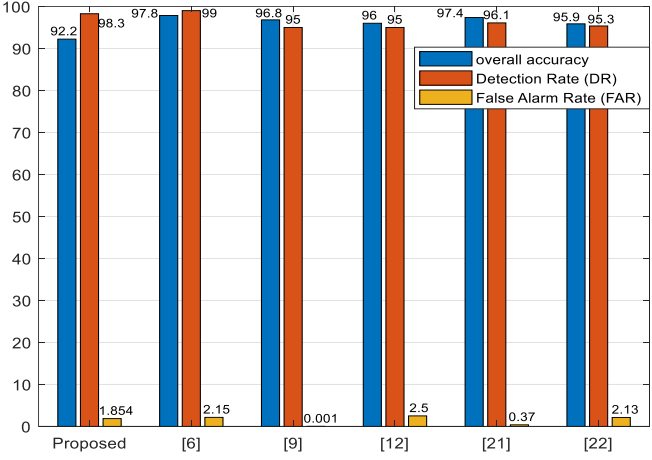

Fig. 4. LGTIDS security evaluation with the metrics of overall accuracy, Detection Rate (DR), and False Alarm Rate (FAR)

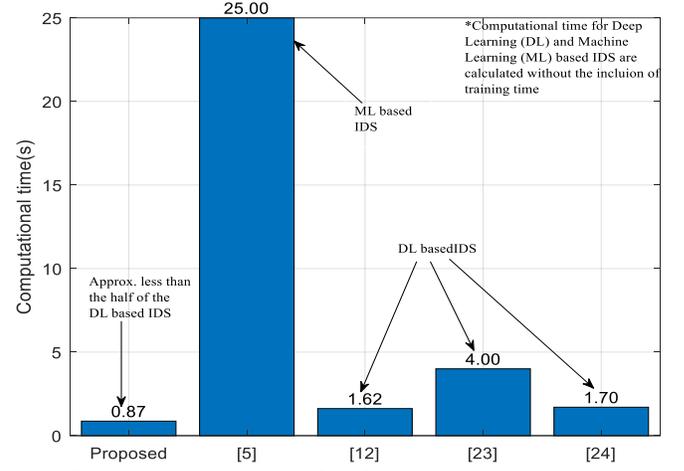

Fig. 5. Computational time analysis for LGTBIDS

TABLE V
PERFORMANCE EVALUATION OF LGTBIDS

| Attack type | Acc.(%) | Precision(%) | Error Rate(%) | FAR(%) | FNR(%) | FOR(%) | Sensitivity(%) | Balanced Acc. (%) | F1 Score(%) | Specificity (%) |
|---|---|---|---|---|---|---|---|---|---|---|
| Half duplex attack | 96.52 | 97.22 | 3.47 | 2.81 | 4.10 | 4.16 | 95.89 | 96.53 | 96.55 | 97.18 |
| Hand over attack | 90.27 | 97.22 | 9.72 | 3.22 | 14.63 | 16.66 | 85.36 | 91.07 | 90.90 | 96.77 |
| Bandwidth spoofing attack | 84.72 | 96.87 | 15.27 | 3.22 | 24.39 | 25 | 75.60 | 86.19 | 84.93 | 96.77 |
| DoS (Denial of Service) | 93.05 | 99.9 | 6.94 | 0.001 | 12.50 | 13.51 | 87.50 | 93.75 | 93.33 | 99.9 |
| UAV attack | 96.52 | 99.9 | 3.47 | 0.001 | 6.49 | 6.94 | 93.50 | 96.75 | 96.64 | 99.9 |

*B. Security*

The security of the intrusion detection mechanism is evaluated by the metrics of True Positive ($T^+$), False Positive ($F^+$), True Negative ($T^-$), and False Negative ($F^-$). $T^+$ is defined as the correct identification of an attack activity or an attack, $F^+$ is the incorrect identification of the legitimate user as an attacker or the incorrect attack alarm alerts for a legitimate network, $T^-$ states the correct identification of the legitimate nodes without malicious activity, and $F^-$ is the incorrect recognition of illegitimate behavior as legitimate. Using these measurements in our simulation, the performance of the proposed mechanism is evaluated.

Fig. 4 shows the comparison of the experimental results contrary to the other mechanisms. The comparison shows that the overall accuracy of the proposed algorithm is higher than the schemes involving machine learning and deep learning mechanism. The overall accuracy of the LGTBIDS algorithm is observed to be 92.22%. Also, the detection rate of the proposed scheme is considerably higher with a specific achieved value of 98.26%. However, the major impact of the proposed scheme is observed on the parameter of FAR. The FAR with a value of 1.853 % is comparatively much lower than the achieved FAR in [6], [12], and [22]. Yet for [9] and [21], the FAR is lower than the proposed mechanism. However, in both cases, the detection rate is comparatively lower than LGTBIDS.

*C. Performance Evaluation*

In the wireless communication environment, intrusion detection systems are typically assessed by real-time performance analysis. The simulation is conducted using five different attacking methods. From the analysis of the results presented in Table V, we observe the overall effectiveness and accuracy of the proposed algorithm. Accuracy is an imperative consideration as far as the security of the WCN is concerned. Table V represents that the proposed model maintains a high range of accuracy ranging from 84.72% to 96.52% with an overall accuracy of 92.2%

The other parameters of the intrusion detection model offer promising effectiveness by maintaining a high level of sensitivity (87.5%), precision (98.2%), F1 score (92.4%), Error rate (7.7%), FAR (1.85%), and specificity (98.1%). Sensitivity outlines 87.5% vertices under attack were correctly detected by LGTBIDS. Specificity identifies that 98.1% of valid vertices are correctly detected. From the examination of the attained results, it is observed that the performance of LGTBIDS for the DoS (Denial of service), UAV (Unmanned Aerial Vehicles), Half-Duplex attack [25], [26] is better than the bandwidth spoofing [11] and hand over attack [28]. It is because of the fact the bandwidth spoofing attack and hand over attack provides a greater variation in the achieved secrecy rate for the distance greater than 170m. Below this distance, the impact of the attack on the achieved secrecy rate is lower. The formula for the performance parameters is given in the appendix. ∎

Hence, it is evident from the obtained results that the proposed IDS algorithm provides a promising performance and provides support in the security of the next generation WCNs. In the WCN, the computational time is characteristically examined as one of the prime parameters that decide the performance evaluation of the IDS algorithms. Based on the layer-wise approach, the nodes from each layer are analyzed simultaneously (in parallel) to improve the computational time. The LGTBIDS mechanism is devoid of the training time, high-end processing, and high-end memory database.

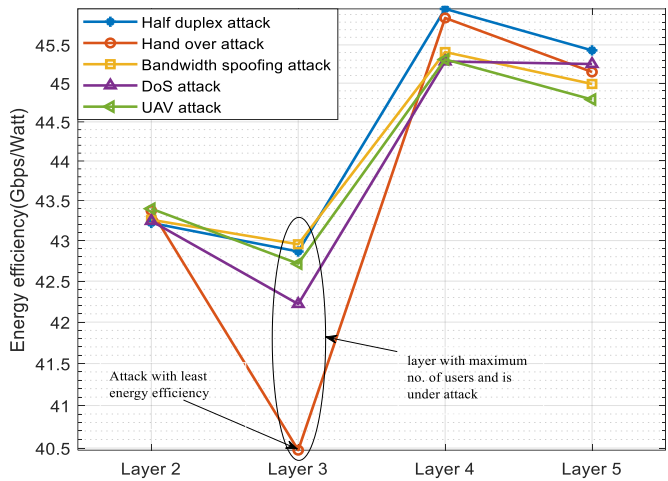

Fig. 6. Energy efficiency analysis LGTBIDS for different attacks.

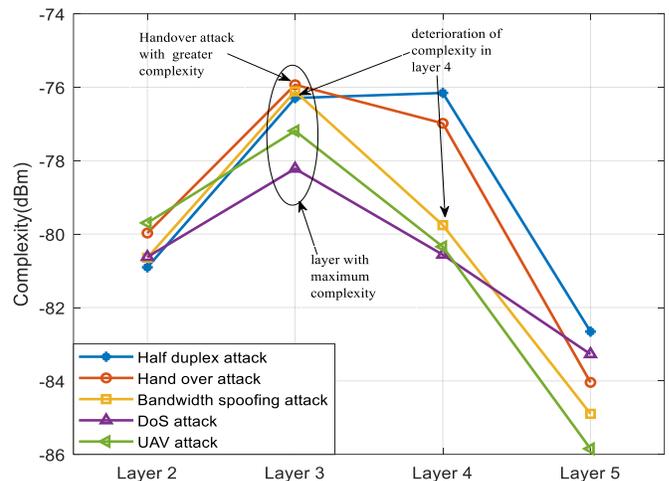

Fig. 7. Complexity evaluation for different attack using LGTBIDS

From Fig. 5. the comparison shows the lowest average computational time of 0.87s (approx.) contrary to the listed intrusion detection schemes. It is because of the fact that only specified vulnerable nodes obtained from the phase of preliminary screening are investigated for the procedure of intrusion detection irrespective of the other number of nodes present in the network. Moreover, layers are simultaneously analyzed to detect the nonstandard node from the network, rapidly and accurately.

Fig. 6. Illustrates the energy efficiency analysis of the LGTBIDS scheme for various attacks. Observing the layer-wise procedure, it is evident that layer 3 is under attack and occupies the maximum number of users. Therefore, the energy efficiency of the IDS in layer 3 is comparatively less significant than in the other layers of the communication network.

Fig. 7. provides insight into complexity parameters for the LGTBIDS approach. The attack is initiated at layer 3 along with the maximum number of users. Therefore, the complexity of layer 3 is higher contrary to other corresponding layers. At layer 3 the hand-over attack is observed to have maximum complexity while incorporating LGTBIDS. On the other hand, the bandwidth spoofing creates a drastic decrease in complexity from layer 3 to layer 4. This is due to the fact that an increase in distance and the multi-stage network favors the bandwidth spoofing attack thereby the difference in intrusion detection becomes quite evident. In the case of a half-duplex attack, the complexity gradually increases from layer 3 to layer 4, because in layer 4 the path loss increases with distance, and the impact of the half-duplex attack becomes uniform which ultimately decreases the prominence effect of the attack thereby increasing the complexity of intrusion detection.

From the overall analysis, the LGTBIDS provides an accuracy of 92.2, a detection rate of 98.3, and a FAR of 1.85. The computational time of 0.87s is maintained by the scheme. Considering FAR, computational time, detection rate, accuracy, and complexity as the prime parameters of the intrusion detection system. The proposed mechanism maintains an efficient balance between the parameters.

It is apparent from the results that the proposed mechanism LGTBIDS offers a validated performance for the procedure of intrusion detection. Also, the mechanism is devoid of the labeled data sets, high processing, and memory end requirements. Moreover, adaptability is ensured by estimating the secrecy rate for the respective vulnerable node with the inclusion of dependent parameters such as distance. Therefore, the proposed LGTBIDS proves an efficient intrusion detection mechanism and identifies a new research direction in the field of security.

*D. Applications*

The proposed mechanism can prove effective in the applications of D2D (Device to Device) communication, spectrum sharing, ultra dense network, internet of things. Numerous other applications such as defense security mechanisms including radar communication systems, UAV, remotely pilot craft, and medical health care can be improved by the proposed security solution.

## VI. CONCLUSION

A graph theory-based Intrusion Detection System is addressed in this paper. The layer-wise analysis is performed such that each layer is scanned for the vulnerable nodes. The approach exploits the characteristics of the signal strength of a node. The efficiency of the algorithm is defined by the parameters of complexity and time computation. The algorithm incorporates each node as the vertex and each transmission channel as the edge of the network. The algorithm involves four phases. The first phases define the network using graph theory. The second phase determines the number of vulnerable nodes. The third phase provides the detection of the attacked node(s) and the fourth phase involves the process of silencing and safeguarding the detected node. The layer-based approach provides the Layer-wise Graph Theory-Based Intrusion Detection System (LGTBIDS) The proposed technique can be further extended to protection mechanisms. Where the transmission of the random sequence is transmitted followed by the search of the communication link such that the secrecy rate lies between the upper bound and lower bound.

APPENDIX

■ *Lemma 1:* The constraint of upper bound limit and lower bound limit in terms of secrecy rate and energy efficiency exist on each of the vulnerable vertices obtained from equation (15) for all $l$ layers, to optimally detect and identify the intruder and valid vertices. The criterion to be fulfilled by the valid user is

mentioned as

i. $(V_s^{ach})_{i,m}^{j,k} \geq (V_s^c)_{i,m}^{j,k}$
ii. $(EE^{ach})_{i,m}^{j,k} \geq (EE^c)_{i,m}^{j,k}$
iii. $(V_s^{ach})_{i,m}^{j,k} \leq (V_s)_{i,m}^{j,k}$
iv. $(EE^{ach})_{i,m}^{j,k} \leq EE_{i,m}^{j,k}$
v. $(V_s^c)_{i,m}^{j,k} \leq (V_s^{ach})_{i,m}^{j,k} \leq (V_s)_{i,m}^{j,k}$
   $(EE^c)_{i,m}^{j,k} \leq (EE^{ach})_{i,m}^{j,k} \leq EE_{i,m}^{j,k}$
vi. Range cross detection function for intrusion detection

$$f(x) = \begin{cases} 0 & f(V_s^{ach}) \in f(V), f(EE^{ach}) \in f(EE) \\ 1 & otherwise \end{cases}$$

Where $x = intruder\ detection$, $V$ lies between $V_s^c$ to $V_s$ and $EE$ lies $EE^c$ to $EE$. 0 denotes that the respective vertex is a valid user and 1 denotes that the corresponding vertex is under attack.

*Proof* To prove *i.*, we note from equation (6) that
$$(V_s^{ach})_{i,m}^{j,k} = |(E^{ach})_{i,m}^{j,k} - E_{i,m}^e| \quad (E^{ach})_{i,m}^{j,k} \geq E_{i,m}^e \quad (30)$$

Using equation (4) in the above equation (30), we get
$$(V_s^{ach})_{i,m}^{j,k} = |(\beta \log_2(1 + \rho_{i,m}^{j,k})) - E_{i,m}^e| \quad (E^{ach})_{i,m}^{j,k} \geq E_{i,m}^e \quad (31)$$

Rearranging equation (31), we get
$$\frac{(V_s^{ach})_{i,m}^{j,k} + E_{i,m}^e}{\beta} = \log_2(1 + (\rho^{ach})_{i,m}^{j,k}) \quad (32)$$

Using equation (2) in equation (32), we obtain
$$\frac{(V_s^{ach})_{i,m}^{j,k} + E_{i,m}^e}{\beta} = \log_2\left(1 + \frac{P_t^{j,k}(i,m)|(h^{ach})_{i,m}^{j,k}|^2}{(\sigma_p^2)_{i,m}^{j,k}}\right) \quad (33)$$

Using $|(h^{ach})_{i,m}^{j,k}|^2 = 10^{-\frac{(p_l^{ach})_{i,m}^{j,k}}{10}}$ and $\frac{P_t^{j,k}(i,m)}{(\sigma_p^2)_{i,m}^{j,k}} = K$, in equation (33), we further obtain

$$\frac{(V_s^{ach})_{i,m}^{j,k} + E_{i,m}^e}{\beta} = \log_2\left(1 + K10^{-\frac{(p_l^{ach})_{i,m}^{j,k}}{10}}\right) \quad (34)$$

Similarly, the lower bound secrecy rate using equation (34) is given by
$$\frac{(V_s^c)_{i,m}^{j,k} + E_{i,m}^e}{\beta} = \log_2\left(1 + K10^{-\frac{(p_l)_{i,m}^{j,k}}{10}}\right) \quad (35)$$

While incorporating additional propagation loss, using equation (10), $(p_l^c)_{i,m}^{j,k} = (p_l^{ach})_{i,m}^{j,k} + p_e$ in the above equation (35), we get
$$\frac{(V_s^c)_{i,m}^{j,k} + E_{i,m}^e}{\beta} = \log_2\left(1 + K10^{-\frac{(p_l^{ach})_{i,m}^{j,k} + p_e}{10}}\right) \quad (36)$$

Comparing equations (34) and (36), we get
$$(V_s^{ach})_{i,m}^{j,k} \geq (V_s^c)_{i,m}^{j,k} \ \forall \ p_e \geq 0 \quad (37)$$

**Part *ii.*** follows part *i*, using equations (7) and (4) the achieved energy efficiency is given as
$$EE_{i,m}^{j,k} = \frac{\beta \log_2(1 + \rho_{i,m}^{j,k})}{P_{cc}} \quad (38)$$

Rearranging equation (38) and using equation (2) in (38) with $|(h^{ach})_{i,m}^{j,k}|^2 = 10^{-\frac{(p_l^{ach})_{i,m}^{j,k}}{10}}$, $\frac{\beta}{P_{cc}} = D$ and $\frac{P_t^{j,k}(i,m)}{(\sigma_p^2)_{i,m}^{j,k}} = K$, we get

$$EE_{i,m}^{j,k} = D\log_2\left(1 + K10^{-\frac{(p_l^{ach})_{i,m}^{j,k}}{10}}\right) \quad (39)$$

Similarly, lower bound energy efficiency using equation (39) with the inclusion of the possible additional propagation loss, using equation (10), $(p_l^c)_{i,m}^{j,k} = (p_l^{ach})_{i,m}^{j,k} + p_e$ is given by

$$(EE^c)_{i,m}^{j,k} = D\log_2\left(1 + K10^{-\frac{(p_l^{ach})_{i,m}^{j,k} + p_e}{10}}\right) \quad (40)$$

Comparing equations (39) and (40), we get
$$(EE^{ach})_{i,m}^{j,k} \geq (EE^c)_{i,m}^{j,k} \ \forall \ p_e \geq 0 \quad (41)$$

From equation (39), we also observe that more increase in the positive value of $p_l$ less is the energy efficiency.

**Part *iii.*** specifies that for a given distance, transmission power the achieved secrecy rate of a valid vertex cannot exceed the maximum constraint (upper bound) of secrecy rate achieved without additional propagation loss given by
$$(V_s^{ach})_{i,m}^{j,k} \leq (V_s)_{i,m}^{j,k} \quad (42)$$

Using the supposition method, suppose the condition given in equation (42) is not true therefore we note the equation as
$$(V_s^{ach})_{i,m}^{j,k} > (V_s)_{i,m}^{j,k} \quad (43)$$

Using equation (6) in equation (43) we get
$$((E^{ach})_{i,m}^{j,k} - E_{i,m}^e) > (E)_{i,m}^{j,k} - E_{i,m}^e \quad (44)$$

Rearranging equation (44), we further obtain
$$(E^{ach})_{i,m}^{j,k} > (E)_{i,m}^{j,k} \quad (45)$$

The equation (45) cannot be true, as the achieved capacity cannot increase beyond Shannon's capacity i.e; capacity achieved without additional propagation loss given in equation (4). The other true condition for equation (45) is given by
$$(E^{ach})_{i,m}^{j,k} \leq (E)_{i,m}^{j,k} \quad (46)$$

Hence our supposition is wrong. Therefore equation (43) cannot be true and the other possible condition is stated as
$$(V_s^{ach})_{i,m}^{j,k} \leq (V_s)_{i,m}^{j,k} \quad (47)$$

**Part *iv.*** follows a similar analogy as given in part *iii*. Multiplying both sides of the equation (46) by $\frac{1}{P_{cc}}$ we get
$$\frac{(E^{ach})_{i,m}^{j,k}}{P_{cc}} \leq \frac{(E)_{i,m}^{j,k}}{P_{cc}} \quad (48)$$

Using equation (7) in the above equation (48) we get
$$(EE^{ach})_{i,m}^{j,k} \leq (EE)_{i,m}^{j,k} \quad (49)$$

**Part *v.*** is addressed by using above stated conditions. Using equations (37) and (42) we can achieve the range of the achieved secrecy rate as
$$(V_s^c)_{i,m}^{j,k} \leq (V_s^{ach})_{i,m}^{j,k} \leq (V_s)_{i,m}^{j,k} \quad (50)$$

Similarly, using equations (41) and (49), we can obtain the constraint of energy efficiency as
$$(EE^c)_{i,m}^{j,k} \leq (EE^{ach})_{i,m}^{j,k} \leq EE_{i,m}^{j,k} \quad (51)$$

**Part *vi.*** addresses the final evaluation step of intrusion detection. Using equations (50) and (51), the comparison is made between the matrix of the possible vulnerable nodes and the criteria stated in equations (50) and (51). The range cross detection function incorporates a one-hot encoding mechanism where 1 is defined for the class of intruders and 0 specifies the valid user. The range cross detection function is stated as

$$f(x) = \begin{cases} 0 & f(V_s^{ach}) \in f(V), f(EE^{ach}) \in f(EE) \\ 1 & otherwise \end{cases} \quad (52)$$

Where $x = intruder\ detection$, $V$ lies between $V_s^c$ to $V_s$ and $EE$ lies $EE^c$ to $EE$. The function compares the achieved secrecy rate and the achieved energy efficiency with the range from lower bounds to upper bounds. If the achieved value of the secrecy rate and the energy efficiency of the corresponding user lies within the range of their upper and lower bounds then the user is considered the valid user. In case, if the achieved value of the secrecy rate and the energy efficiency falls beyond the range of upper and lower bounds of the respective secrecy rate and energy efficiency then the user is declared as the user under attack.

■ The parameters of the performance evaluation are given as

$$Accuracy = \frac{T^+ + T^-}{T^+ + T^- + F^+ + F^-} \quad (53)$$

$$Detection\ rate = \frac{T^+}{T^+ + F^+} \quad (54)$$

$$False\ Alarm\ Rate\ (FAR) = \frac{F^+}{F^+ + T^-} \quad (55)$$

$$False\ Omission\ Rate\ (FOR) = \frac{F^-}{F^- + T^-} \quad (56)$$

$$False\ Negative\ Rate\ (FNR) = \frac{F^-}{T^+ + F^-} \quad (57)$$

$$Specificity = \frac{T^-}{F^+ + T^-} \quad (58)$$

$$Sensitivity = \frac{T^+}{T^+ + F^-} \quad (59)$$

$$Balanced\ Accuracy = \frac{Sensitivity + Specificity}{2} \quad (60)$$

$$F1\ score = \frac{2T^+}{2T^+ + F^+ + F^-} \quad (61)$$

$$Error\ rate = \frac{F^+ + F^-}{T^+ + T^- + F^+ + F^-} \quad (62)$$

ACKNOWLEDGMENT

The authors gratefully acknowledge the support provided by 5G and IoT Lab, DoECE, and TBIC, Shri Mata Vaishno Devi University, Katra, Jammu.